\documentclass{iau}
\usepackage{graphicx}
\usepackage{amsmath}

\title[Parallelization of the SIR code] 
{Parallelization of the SIR code for the investigation of small-scale features in the solar photosphere}

\author[S. Thonhofer et al.]   
{Stefan Thonhofer$^{1,2}$,
Luis R. Bellot Rubio$^2$,
Dominik Utz$^{1,2}$,\\
Arnold Hanslmeier$^1$,
Jan Jur\v c\'ak$^3$
 }

\affiliation{$^1$Institute of Physics, University of Graz \\
Universit\"atsplatz 5, A-8010 Graz, Austria \\email: {\tt stefan.thonhofer@edu.uni-graz.at} \\[\affilskip]
$^2$Instituto de Astrof\'isica de Andaluc\'ia, CSIC \\
Glorieta de la Astronom\'ia s/n, E-18008 Granada, Spain \\[\affilskip]
$^3$Astronomical Institute of the Academy of Sciences \\
Fri\v cova 298, CZ-25165 Ond\v rejov, Czech Republic}

\pubyear{2015}
\volume{305}  
\jname{Polarimetry: From the Sun to Stars and Stellar Environments}
\editors{K.N. Nagendra, S. Bagnulo, \\ R. Centeno, \& M. Mart\'inez 
Gonz\'alez, eds.}
\begin{document}

\maketitle

\begin{abstract}
  Magnetic fields are one of the most important drivers of the highly
  dynamic processes that occur in the lower solar atmosphere. They
  span a broad range of sizes, from large- and intermediate-scale
  structures such as sunspots, pores and magnetic knots, down to the
  smallest magnetic elements observable with current telescopes. On
  small scales, magnetic flux tubes are often visible as Magnetic
  Bright Points (MBPs).  Apart from simple $V/I$ magnetograms, the
  most common method to deduce their magnetic properties is the
  inversion of spectropolarimetric data. Here we employ the SIR code
  for that purpose. SIR is a well-established tool that can derive not
  only the magnetic field vector and other atmospheric parameters
  (e.g., temperature, line-of-sight velocity), but also their
  stratifications with height, effectively producing 3-dimensional
  models of the lower solar atmosphere.  In order to enhance the
  runtime performance and the usability of SIR we parallelized the
  existing code and standardized the input and output formats. This
  and other improvements make it feasible to invert extensive
  high-resolution data sets within a reasonable amount of computing
  time. An evaluation of the speedup of the parallel SIR code shows a
  substantial improvement in runtime.  
\end{abstract}
\firstsection

\section{Introduction}
For the investigation and understanding of processes involving
small-scale magnetic features in the solar photosphere it is essential
to obtain detailed information about the underlying magnetic field.
This can be done through simulations or observations. In the case of
observations one can choose among several possibilities to study the
properties of the field. Proxymagnetometry, for instance, exploits the
intensity variations caused by the magnetic field to detect magnetic
features and follow their evolution (\cite{steiner} or
\cite{schussler}).  Other methods use (spectro)polarimetry: due to
the Zeeman effect, magnetic fields induce polarization in spectral
lines and thus spectropolarimetric observations can yield insights
into the physics of the solar atmosphere. For the interpretation of
the data one can apply sophisticated computer codes which solve the
radiative transfer equation and determine the physical conditions
under which the light was emitted. Our group uses the SIR code (Stokes
Inversion based on Response Functions, see \cite{sir1992} and
\cite{lbr}) for the inversion of spectropolarimetric measurements.
SIR retrieves not only the atmospheric parameters (temperature $T$,
line-of-sight velocity $v_{LOS}$, magnetic field strength $B$, and
magnetic field angles), but also their stratifications along the line
of sight. This is done by computing synthetic Stokes spectra from an
initial guess model, comparing them to the observations, and modifying
the model until the synthetic profiles fit the observed ones (i.e., the
$\chi^2$ values become minimum). For this fitting procedure a
Marquardt algorithm is used.  Sometimes the algorithm is unable to
find the global minimum of the $\chi^2$ function and the code fails to
converge.

The original SIR code can only invert one pixel at a time, while
modern solar polarimeters provide us with an enormous wealth of data
consisting of millions of pixels within a time series. For the
inversion of such huge amounts of data we parallelized the SIR code
(see Section \ref{sec:par}). We also introduced a new input/output
strategy (Section \ref{sec:io}) and implemented the possibility of
using multiple initial model atmospheres to solve rare cases of
non-convergence (Section \ref{sec:multinit}). The achieved speedup 
is discussed in Section \ref{sec:speedup} and the conclusions are
summarized in Section \ref{sec:conclusion}.

\section{Parallelization and extension of the SIR code}

\subsection{Parallelization Strategy}
\label{sec:par}
\begin{figure}[t]
\begin{center}
\includegraphics[width=0.7\textwidth]{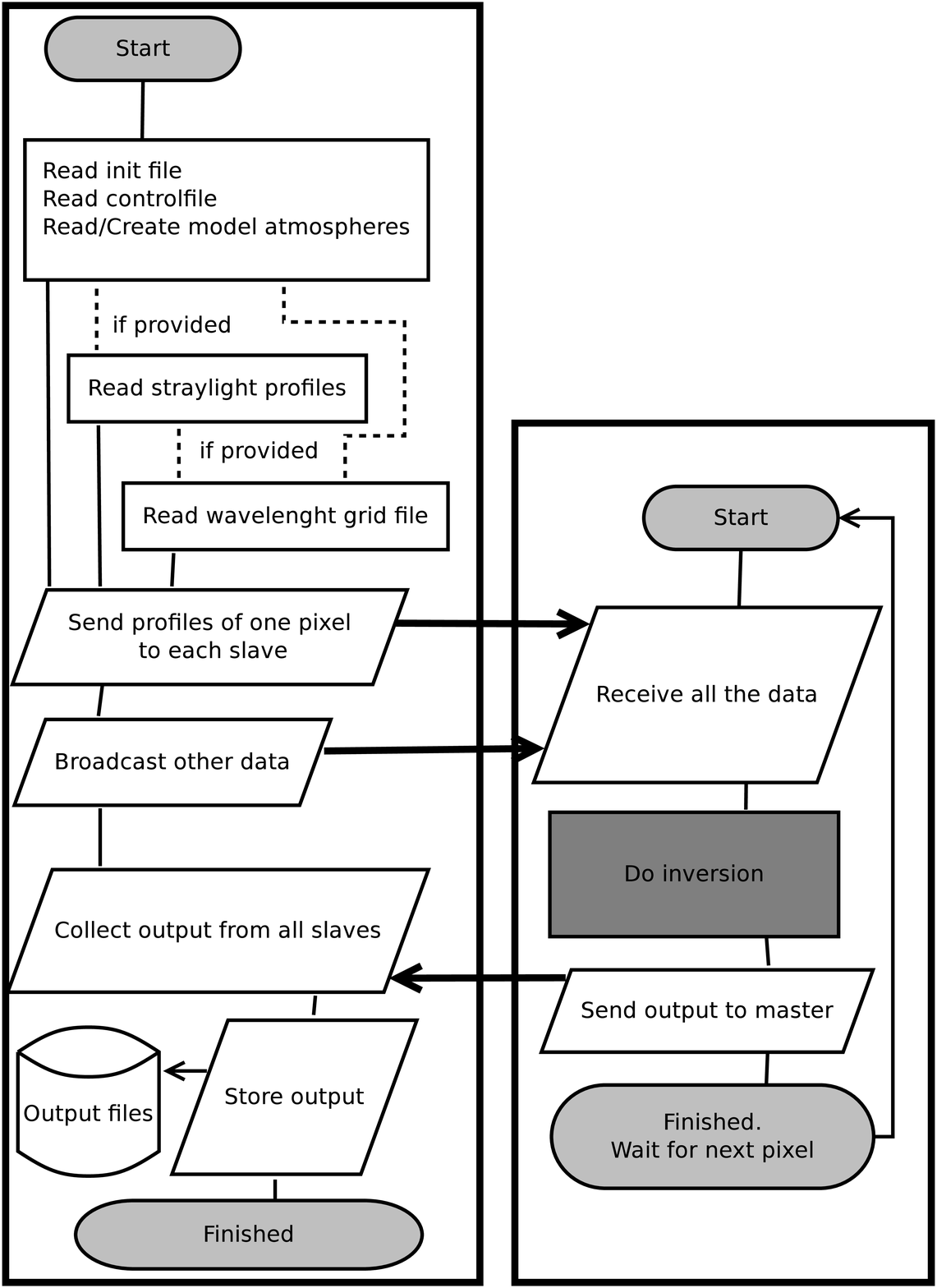}
\caption{Flow chart of the parallel SIR code. The procedure tree
  inside the left box corresponds to the master process, whereas the
  tree within the right box represents one of the slave processes. The
  module labeled ``Do inversion'' essentially contains the
  functionality of the serial SIR code.}
\label{fig:flow}
\end{center}
\end{figure}

We included MPI (Message Passing Interface) routines in the SIR code,
enabling the usage of a cluster computing system for the inversion.
Figure \ref{fig:flow} shows the workflow of the parallel SIR code. A
master-slave topology was incorporated, i.e., one master process
performs all input/output and distributes the work to slave processes.
This is possible since the pixels are treated separately and inverted
independently of each other. In order to achieve a high speedup, we
had to modify the input and output modules of SIR (see next section).
Except for that, the functionality and numerical setup of the original
SIR code remains the same.

\subsection{New input/output features}
\label{sec:io}
While the serial SIR code uses ASCII input and output files for each
single pixel, we had to implement a more sophisticated format to
better handle large amounts of data. We chose the FITS (Flexible Image
Transport System, see \cite{fits}) standard as it is widely spread
among the scientific community and easy to use.  The following
(partially optional) changes regarding the input/output were made
compared to the serial version of the SIR code:
\begin{itemize}
\item Input of the observed Stokes spectra via FITS files.
\item Optional input of a pixel mask for the inclusion and exclusion
  of single pixels.
\item Optional inversion of a time sequence using the same settings
  for each time step.
\item Optional use of a different wavelength grid and stray light
  profile for each pixel.
\item Output of all inversion results\footnote{synthetic spectra,
    retrieved model atmospheres and their errors, plus additional
    information} via FITS files.
\end{itemize}

We introduced a so-called \texttt{init} file enabling the user to
control the new features of the parallel SIR code. The control file of
the original SIR code remains the same and still contains the main
settings of the inversion, such as the names of the atomic data and
abundance files, the atmospheric parameters to be retrieved, and the
number of nodes for each physical quantity.

The current state of the inversion (number of inverted pixels, number
of pixels to be inverted, elapsed time, estimated time of the whole
inversion) is written to a text file called \texttt{PROGRESS} to
inform the user about the inversion progress.

To prevent data loss in case of the failure of a process, the
results are periodically written to the output files. The frequency of
this backup is specified in the \texttt{init} file.

\subsection{Multiple initial model atmospheres}
\label{sec:multinit}
For certain combinations of input data (spectra, initial models), the
SIR code may fail to converge. Moreover, experience shows that in
regions with weak polarization signals the solution may depend on the
initial guess model. In order to avoid non-convergence and improve the
best-fit $\chi^2$-values, we implemented the possibility of using
multiple initializations. If this option is selected, a set of
different initial model atmospheres is created in a permutative way
and the inversion of each pixel is repeated using all these models. In
the end, the solution providing the best fit (i.e., the lowest
$\chi^2$) is selected and used for the output.

The values of the initial models are defined by the user and should
cover the parameter space as densely as possible. In particular, the
initial gradients of $B$, $v_{LOS}$, and field inclination can be
specified, both in sign and in magnitude. Trying different
combinations of the parameters increases the reliability of the
inferences, especially when it comes to determining the sign of the
gradients from noisy and/or coarsely sampled Stokes profiles

Tests have shown that the code often converges very well with just one
initialization. The advantage of different initializations is that
they resolve rare cases of non-convergence and also ensure that the
correct solution is chosen among weak and strong fields, large and
small inclinations to the vertical, etc. Thus, they usually lead to
better $\chi^2$-values.

\section{Speedup}
\label{sec:speedup}
\begin{figure}[t]
\begin{center}
\includegraphics[width=0.8\textwidth]{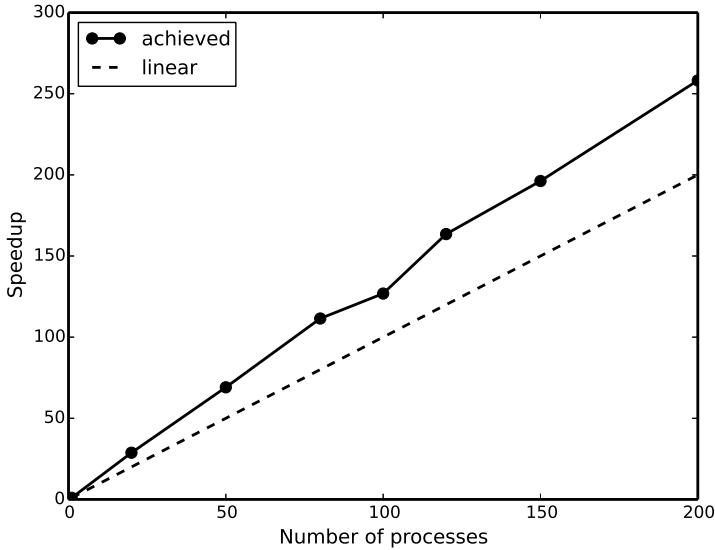}
\caption{Speedup of the parallel SIR code with respect to the serial
  version. The dashed line represents linear speedup.}
\label{fig:speedup}
\end{center}
\end{figure}

The quality of a parallelization can be measured in terms of the
speedup, $S_N$, defined as
$$
S_N=\dfrac{t_1}{t_N},
$$
where $t_1$ is the execution time for only one processor (in our case,
running the serial version of the SIR code), and $t_N$ is the
execution time using $N$ processors. In the ideal case, the
relationship is linear, i.e., $S_N=N$. In practice, however, the
speedup is limited by non-parallelizable operations, such as file
input/output or memory access.  Figure \ref{fig:speedup} displays the
speedup of the parallel SIR code. We measured the execution times for
the same pixels and settings using a different number of MPI processes
on the cluster system of the Instituto de Astrof\'isica de
Andaluc\'ia. In order to obtain the reference value $t_1$ we applied
the serial SIR code to the same pixels.

It is remarkable that the speedup is superlinear\footnote{We wish to
  note that the actual outcome of this experiment may vary slightly
  depending on the utilization of the cluster by other users.} in each
measurement. This behavior is most probably caused by cache effects:
if $N$ is large enough, the capacity of the accumulated cache will be
sufficient to hold enough data of the program as to reduce the memory
access time, thus leading to an increased speedup. In addition, the
input/output operations are reduced to a minimum in the parallel SIR
code. These operations are very time consuming because they involve
disk access, so avoiding them significantly increases the code
performance.

\section{Conclusion}
\label{sec:conclusion}
SIR is an advanced tool for the numerical inversion of the radiative
transfer equation. It enables the user to deduce physical quantities
within the solar photosphere and their gradients from observed Stokes
spectra. In order to facilitate the application of the code to modern
high-resolution data sets we improved its runtime performance by fully
parallelizing it. This improvement goes along with other new features,
e.g., the implementation of new input/output routines for direct
access to FITS files. The new version of SIR can be run on
high-performance clusters to invert large datasets within a reasonable
time span. The most important improvements with respect to the
original SIR code are:
\begin{itemize}
\item The parallelization enabling the inversion of more than one pixel per call using several processor cores.
\item Reading the input spectra from standardized FITS files.
\item Writing all output data to standardized FITS files.
\item Excluding single pixels from the inversion via the usage of a pixel mask.
\item Inversion of a whole time sequence in one call.
\item Periodic backup of output results in order to prevent data loss.
\item Using different initial model atmospheres to improve the fits.
\end{itemize}
Most of these features are controlled via the newly introduced
\texttt{init} file. Information about the current status of a running
inversion is written to a file called \texttt{PROGRESS}.

The speedup of the parallel SIR code compared to the original version
shows a superlinear trend. This is a remarkable behavior for a
parallel program. Already a linear speedup would indicate a proper
implementation of the parallelization (i.e. no systematic errors were
made during the conceptual design phase of the parallelization as well
as during the programming phase). The superlinear trend is most
probably linked to the significant input/output optimizations and
beneficial cache effects.

The user of the parallel SIR code can optionally apply a new strategy
of inversion. Instead of running the inversion with only one initial
model atmosphere, a set of different models covering the range of
meaningful parameter values can be employed to invert each pixel.  The
solution providing the best fit to the data is selected and stored as
output. First tests have already shown the validity of this approach.

The parallel version of SIR can be downloaded from {\tt
  http://spg.iaa.es/downloads}. In the future, we also plan to
parallelize SIRGAUSS and SIRJUMP (\cite{sirgauss}). These codes
implement Gaussian perturbations and discontinuities of the atmospheric
parameters along the line of sight, respectively, and can be used to
invert spectra from sunspot penumbrae and the quiet Sun. 

\section*{Acknowledgements}
This work was funded by the Austrian Science Fund (FWF): P23618
(Dynamics of Magnetic Bright Points), the LLP ERASMUS program of the
E.C., the F\"or\-der\-ungs\-stip\-end\-ium of the University of Graz,
and the Spanish Ministerio de Econom\'{\i}a y Competitividad through
grants AYA2012-39636-C06-05 and ESP2013-47349-C6-1-R, including a
percentage from European FEDER funds. D.U.  wants to acknowledge the
special support given by project J3176 (Spectroscopic and Statistical
Investigations on MBPs).  J.J. is thankful to the \"OAD
(\"Osterreichischer Austauschdienst) and the M\v{S}MT (Ministry of
Education, Youth and Sports, Czech Republic) for supporting research
stays in Austria. S.T. and D.U. are also thankful to the \"OAD and the
M\u{S}MT for supporting research stays in the Czech Republic.  The
authors gratefully acknowledge support from NAWI Graz.

\end{document}